\documentclass[fleqn,10pt]{wlscirep}
\usepackage{amssymb,amsmath}
\usepackage{color}

\title{Two-color pump-probe interferometry of ultra-fast light-matter interaction}

\author[1*]{Yoshio Hayasaki}
\author[1]{Shin-ichi Fukuda}
\author[1]{Satoshi Hasegawa}
\author[2,3*]{Saulius Juodkazis}

\affil[1]{Center for Optical Research and Education (CORE),
Utsunomiya University, 7-1-2 Yoto, Utsunomiya 321-8585, Japan}
\affil[2]{Centre for Micro-Photonics, Faculty of Science,
Engineering and Technology, Swinburne University of Technology,
Hawthorn, VIC 3122, Australia}\affil[3]{Center of Nanotechnology,
King Abdulaziz University, Jeddah 21589, Saudi Arabia}
\affil[*]{For correspondence: Y.H. hayasaki@cc.utsunomiya-u.ac.jp;
S.J. sjuodkazis@swin.edu.au}


\begin{abstract}
Two-color side-view probing of light-matter interaction from
minute focal volume of a tightly focused fs-laser pump pulse
reveals charge dynamics with high 0.9~$\mu$m optical resolution
and approximately $\sim 20$~fs temporal resolution in coincidence
between pump and probe pulses. Use of two colors is advantageous
for probing optically excited plasma regions with different
density. Holographical digital focusing and spatial filtering were
implemented to obtain the same resolution images for subsequent
Fourier analysis. Fast electron removal with time constant $\sim
150$~fs was resolved and is consistent with self-trapping.
Potential applications of an optical control over a light-induced
defect placement with deep-sub-wavelength resolution is discussed.
\end{abstract}
\begin{document}
\flushbottom \maketitle

\thispagestyle{empty}

\section{Introduction}

Imaging and detection of ultra-fast phenomena is in constant
development using different methods of pump-probe techniques at
different magnifications and
resolutions~\cite{Psaltis,Gong,Tadas,Mu,Solis,Razvan}. High
spatial resolution of a pump-probe optical imaging of tightly
focused fs-laser pulses inside glass carried out in
lateral~\cite{11oe5725} and axial~\cite{11ome1399} views with
interferometric technique was $\sim 0.2, 0.6~\mu$m, respectively.
The side-view (axial) imaging is sought after due to possibility
to use shorter wavelength strobe (probe) pulses and to achieve
high resolution even at moderate focusing with objective lens of
numerical aperture $NA \simeq 0.5$. Formation of plasma filaments,
self-trapped excitons, shock waves, axially extended
voids~\cite{Bhuyan}, tubular compressed micro-volumes~\cite{Xie}
and bulk ripples~\cite{Rudenko} can be potentially resolved using
such side-view imaging. Those phenomenon are highly dynamic and
have transient stages which are not observed in standard
\emph{post mortem} inspection of samples or are detected by with
time integration causing a lost temporal
resolution~\cite{Velpula,11nc445,10prb054113,16le16133}.
Pump-probe methods are fast evolving to improve spatial and
temporal resolutions, e.g., tilted fs-laser pulses were used to
increase region of spatial-temporal overlap in imaging of
fillaments in water~\cite{Tadas}, to reveal the mechanism of water
ionisation~\cite{Audrius}, digital holography was applied to reach
high temporal resolution of filamentation and ablation of
transparent dielectrics~\cite{Siaulys}, the Abel inversion
 was used to reconstruct the temporal density evolution of
plasma in air~\cite{Papazoglou}.

Here, a novel interferometric pump two-color probe method is
developed for a high spatial and temporal imaging of ultra-fast
light-matter interaction in the bulk of transparent sample.
Combination of two-color probes allows to better visualise plasma
excited by the pump pulse due to difference in absorbance.

\begin{figure}[tb]
\begin{center}
\includegraphics[width=9.50cm]{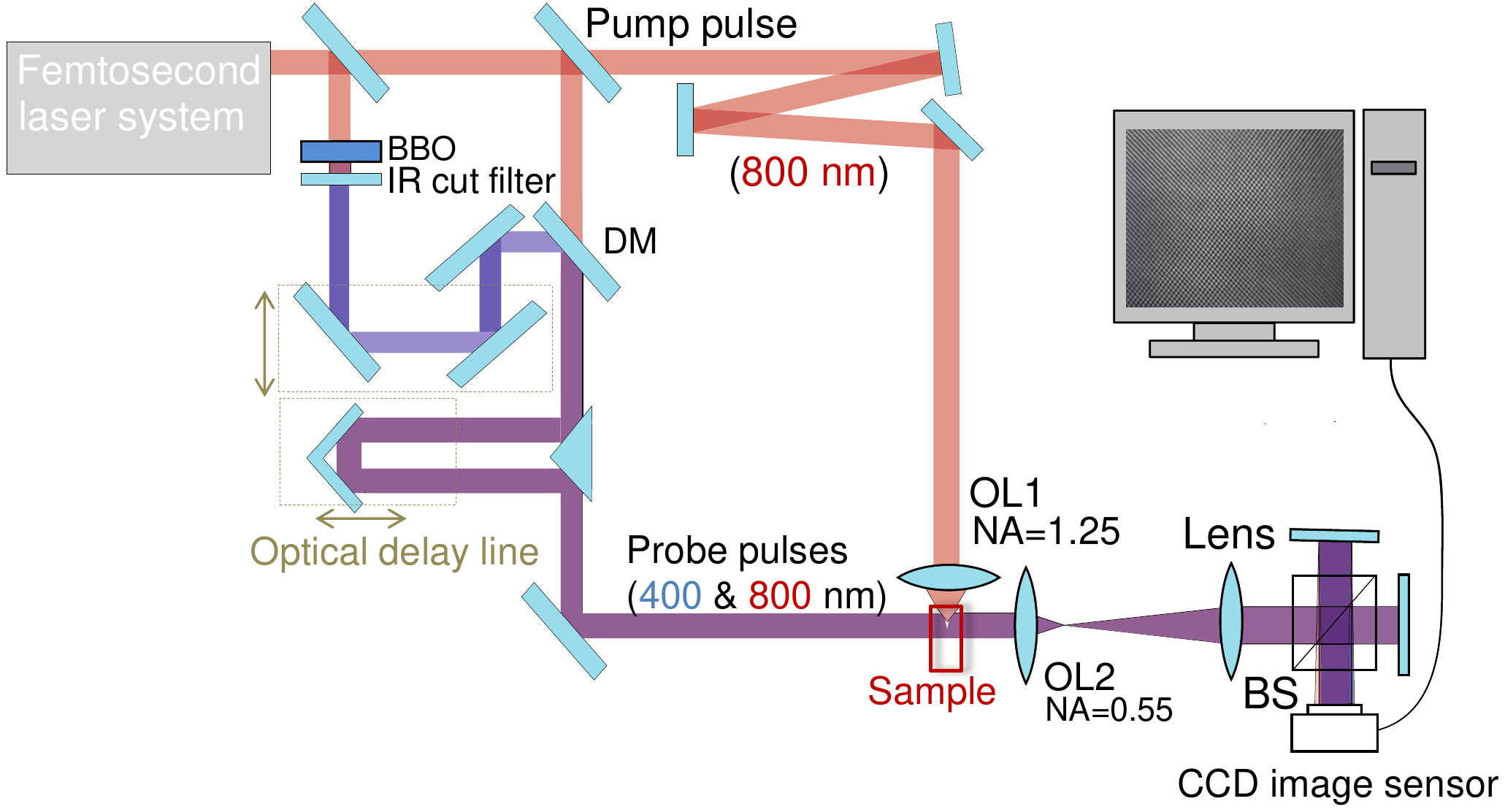}
\caption{Setup of the pump and two color probe interferometry.
OL1, 2 are the objective lenses with corresponding numerical
apertures NA, BS is the beam splitter , DM - dichroic mirror to
combine 800 and 400~nm pulses, BBO - $\beta$BaBO$_4$ crystal is
for frequency doubling ($\lambda = 400$~nm wavelength). }
\label{f-setup}
\end{center}
\end{figure}

\begin{figure}[tb]
\begin{center}
\includegraphics[width=8.0cm]{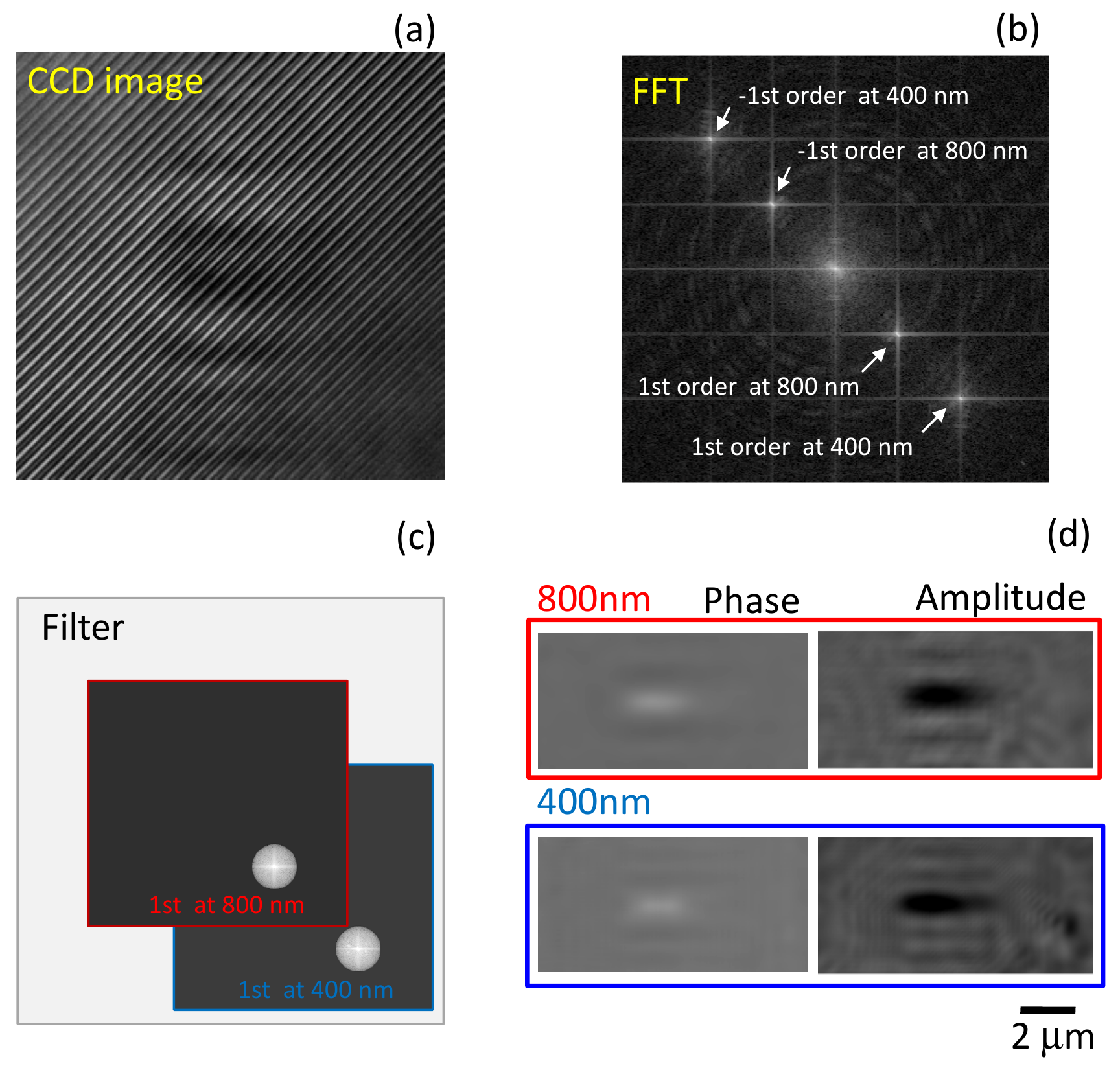}
\caption{Implementation of the method. A CCD image (a) and its
Fourier transform (b) with selection of the filter apertures for
back transform (c) with the resulting phase and amplitude images
(d). The filters  (in (c)) are set to the optical resolution of
setup at 0.89~$\mu$m.} \label{f-method}
\end{center}
\end{figure}

\section{Experimental}

\subsection{Optical setup}

Figure~\ref{f-setup} shows an experimental setup of the pump-probe
interference microscope with two-color probes. The pump pulse with
a center wavelength of $\lambda_\omega = 800$~nm generated by an
amplified femtosecond (fs)-laser system was focused inside a glass
sample from its sidewall with a $100^\times$ oil-immersion
objective lens (OL) with a numerical aperture $NA$ of 1.25. The
sample was a superwhite crown glass (B270, Schott), which is
representative of the widely used low dispersion glasses. The
laser-induced phenomena was induced by a single pulse irradiation
and the sample was moved transversely every one pulse. Focus was
placed only $10~\mu$m below the surface to minimize spherical
aberration.

The two-color probes are the fundamental wavelength of
$\lambda_\omega = 800$~nm and its second-harmonics with a center
wavelength of $\lambda_{2\omega} = 400$~nm. The second-harmonics
probe pulse was generated in a barium borate (BBO) crystal. Both
two-color probe pulses were irradiated onto the sample with a
delay time $T$ after arrival of the pump pulse and was introduced
to the interferometer after passing through the sample using a
telocentric microscope optics composed of an $50^\times$ OL with a
focal length of 4.0~mm and an $NA$ of 0.55 and a doublet lens with
a focal length of 500~mm.

Control of the pump pulse duration was made by an \emph{in situ}
monitoring the breakdown inside the glass and tuning an
outside-cavity compressor: the lowest pulse duration at which the
optical breakdown was observed was approximately 0.2~ps as
measured by the fastest change of the phase at the focal spot. The
pulse duration at the laser output was 45~fs. The duration of the
probe pulse was minimized by monitoring the most efficient SHG
while pre-chirping pulse at the fundamental frequency.

The delay $T$ was controlled with an optical delay line with the
maximum delay of 10~ns. The delay line was composed of two
stepping-motor-driven stages for controlling the fine and coarse
movements. The fine stage for the delay less than 1~ps had a step
movement 0.5~$\mu$m, a resolution of 3~$\mu$m at the maximum
translation of 25~mm. The coarse stage for delays longer than 2~ps
had a translation step of 25~$\mu$m, resolution of 16~$\mu$m, and
the stroke length of 1500~mm.

The coincidence $T = 0.0$~ps time between pump and probe was
decided when there was no recognizable photomodification at the
pump pulse of $\sim$50~nJ; hence, when actual separation between
counter-propagating pump and probe pulses was approximately equal
to the pump pulse duration. Precision of the setting of the time
reference frame was approximately $\pm$20~fs. In this study we
focussed on plasma formation before the void formation which was
observed for the pump pulse energies above 300~nJ at the same
focusing conditions as was reported previously~\cite{11ome1399}.

\section{Image analysis}

Interference fringes were recorded by the angularly multiplexed
method with a cooled charge-coupled device (CCD) image sensor
(BU-50LN, Bitran) with a pixel size of 8.3~$\mu$m and 16-bit
recording, and read out by a computer for analysis. The angularly
multiplexed method described later in detail~\cite{Psaltis}.

The plasma emission was observed at the focusing point, however,
the intensity was not so strong in the used glass. The integrated
intensity of plasma emission was $<$20\% as compared with
intensity of the probe beam with the image sensor detection. The
plasma emission superimposed on the interference images was
subtracted from the raw interference image before numerical
procedures.

Complex amplitude of the interference image was obtained by the
Fourier filtering method. of the interference image~\cite{Takeda}.
More detailed procedure is described below. Let us consider an
interference between an one-dimensional object wave with an
amplitude $a_s(x)$ and a wave number $k_s(x)$, a plane reference
wave with an amplitude $a_r$ with a wave number $k_r$. It is
supposed that $k_s(x)$ has the object signal components
$k_{s1}(x)$ around the center $k_{s0}$. The fringes for each
wavelength is described as:
\begin{equation}\label{e1}
I(x)= a_r^2 + a_s(x)^2 + a_ra_s(x)\exp i[-Kx-\phi(x)] +
 a_ra_s(x)\exp i[Kx+\phi(x)],
\end{equation}
\noindent where $K = k_{s0} - k_r$ and $\phi(x)=k_{sl}(x)x$. The
wave number of the fringes at 400~nm, denoted as $K_2$, is twice
of that for 800~nm, denoted as $K_1$, that is $K_2 = 2K_1$. The
Fourier transform of the superposition of the fringes formed by
two wavelengths described as:
\begin{equation}\label{e2}
F\left[\sum_{n = 1,2}I_n(x)\right] = \sum_{n=1,2}A_n(\omega) +
B_n(\omega+K_n)^* + B_n(\omega-K_n),
\end{equation}
\noindent where $F$ is the Fourier transform, $n = 1$ and $n = 2$
are the values of 800~nm and 400~nm, respectively,
$A_n(\omega)=F[a_{n,r}^2 + a_{n,s}^2(x)]$ and $B_n(\omega-K) =
F[a_{n,r}a_{n,s}(x)\exp(i\phi(x))\exp(iK_nx)]$, and $^*$ marks the
complex conjugate. If $max[A(\omega)] + max[B_1(\omega)]<K_1$ and
$max[B_1(\omega)] + max[B_2(\omega)]<K_1$, that is, the spectra of
$A(\omega)$, $B_1(\omega)$, and $B_2(\omega)$ have no overlap over
each other, they can be separated by a spatial frequency filter on
the Fourier plane as shown in Fig.~\ref{f-method}(b). The filter
$W(\omega)$ used in the experiments was of the Hanning type,
defined by:
\begin{equation}\label{e3}
W(\omega) = \begin{cases} [1+\cos(\mid \omega - K_n\mid/2h)]/2, &
\mathrm{if}
\mid \omega - K_n\mid\leq 2\pi h\\
0, & \mathrm{otherwise}\end{cases}
\end{equation}
\noindent where $h$ is the filter width.

Resolutions of the interference observation were 0.44~$\mu$m at
400~nm and 0.89~$\mu$m at 800~nm, which were calculated from a
full width at half maximum (FWHM) of the Airy disk pattern using
the $NA=0.55$ objective lens; the magnification of the microscope,
$M$, was 125 and the cut-off spatial frequency of the Fourier
filtering $h$ (described in the next section) was 12 line pairs
per millimeter (lp/mm).

In the framework of geometrical optics and with assumption of
aplanatic focusing, a computed cut-off frequency of the total
system was $Mh =1470$~lp/mm and the corresponding resolution limit
was $1/Mh = 0.68~\mu$m. This is larger than the resolution of
$0.44~\mu$m at 400~nm and smaller than the resolution of
$0.89~\mu$m at 800~nm). In order to equalize both resolutions, the
cut-off frequency of the filter for the 400~nm images was
decreased. After both $B_\lambda(\omega-K)$ were extracted by the
spatial frequency filter, the complex amplitudes were obtained by
the inverse Fourier transform as:
\begin{equation}\label{e4}
b(x) = a_ra_s(x)e^{i\phi(x)}e^{iKx}.
\end{equation}
Each measurement had two image captures. First, image was captured
just before the pump-pulse irradiation to obtain the reference.
Second, the image was taken with pump present and the ratio was
calculated. The used procedure was tested experimentally to have a
negligible effect of cross talk between the two filters at 400~nm
and 800~nm. However, for more complex images with high content of
high spatial frequencies the cross talk might occur. It could
still be eliminated by selection of a slightly different angles of
incidence for the two probes.

In order to eliminate the tilt component in the output images,
they were normalized by the output images obtained previously
without an object, described as $b_0(x)=a_ra_ie^{iKx}$, where
$a_i$ is the amplitude of the illumination light. Finally, the
normalized output images were obtained as
\begin{equation}\label{e5}
\widetilde{b}(x)= \frac{a_s(x)}{a_i}e^{i\phi(x)},
\end{equation}
\noindent where $a_s(x)/a_i$ is the transmittance of the object.

\begin{figure}[tb]
\begin{center}
\includegraphics[width=16.50cm]{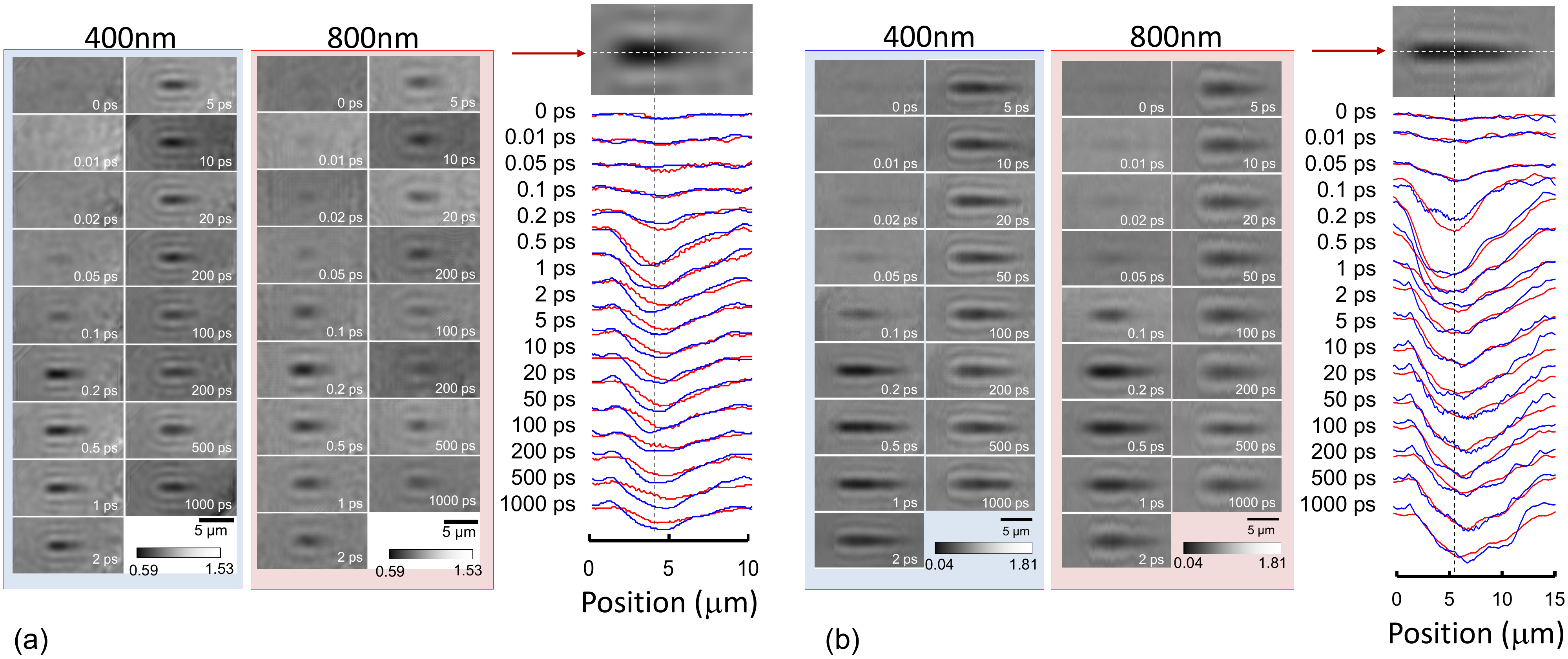}
\caption{Amplitude of FFT image at 400~nm and 800~nm wavelengths
after excitation pulse of $E_p = 50$~nJ at focus (a) and 200~nJ
(b) excitation together with the on-axis cross sections along
pulse propagation; the amplitude scale was the same for both
wavelengths. Measurements at both wavelengths was carried out
simultaneously. } \label{f-ab}
\end{center}
\end{figure}
\begin{figure}[tb]
\begin{center}
\includegraphics[width=13.50cm]{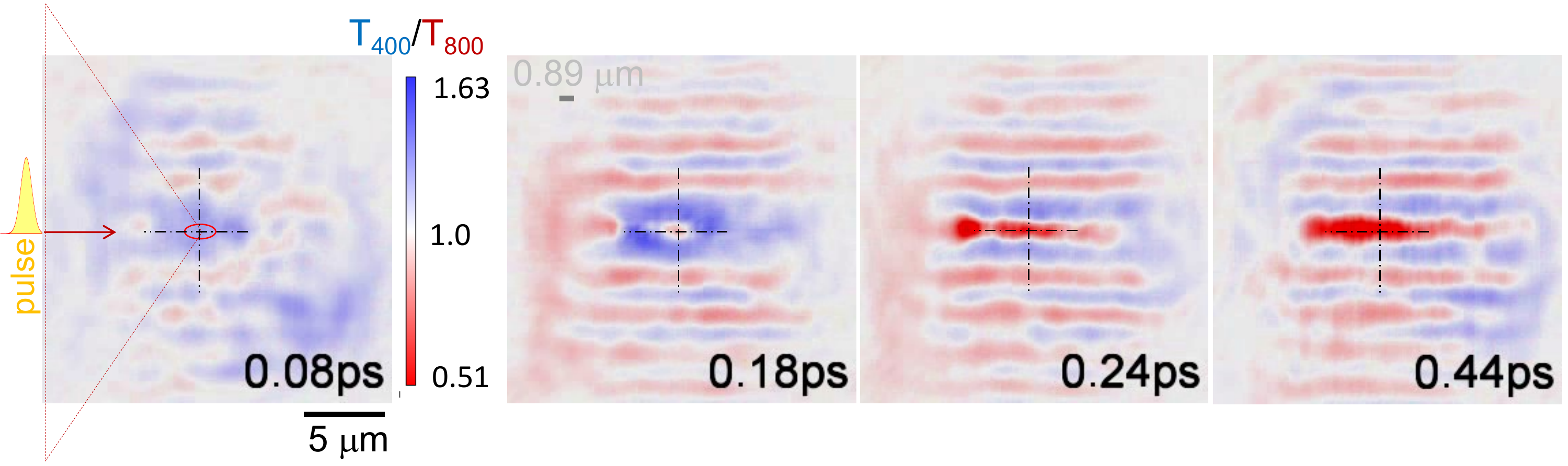}
\caption{Transmission ratios at 400~nm and 800~nm,
$T_{400}/T_{800}$, at different times after single pulse of $E_p =
200$~nJ focused by $NA\equiv n\sin\alpha = 1.25$ ($\alpha =
55.6^\circ$) objective lens (shown by a triangle); the optical
resolution  of the FFT filter was 0.89~$\mu$m marked by ellipse.
Fringing is caused by division of two images with similar
intensity values and the FFT caused oscillations are enhanced. }
\label{f-200}
\end{center}
\end{figure}
\begin{figure}[tb]
\begin{center}
\includegraphics[width=9.50cm]{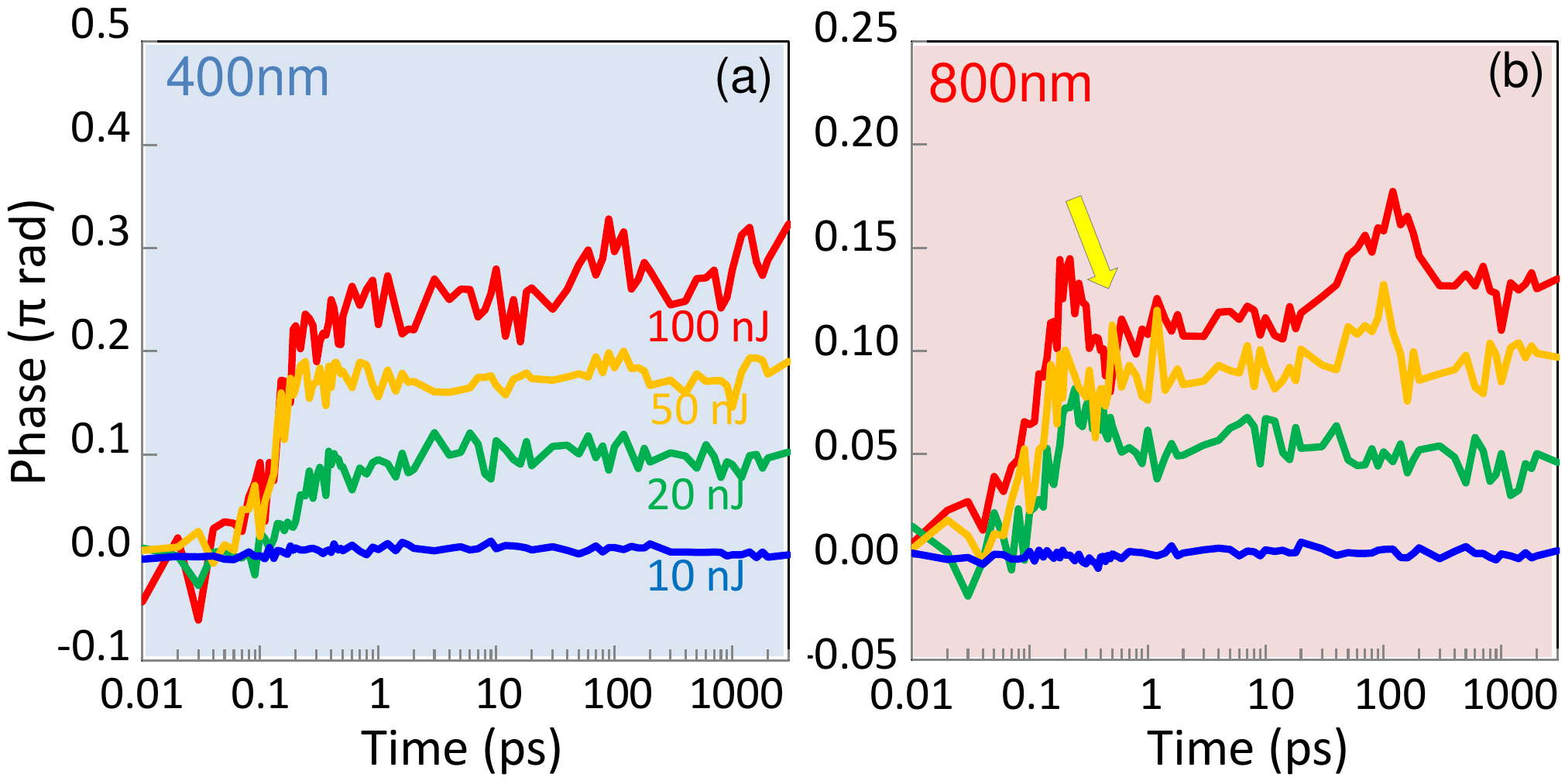}
\caption{Phase of FFT images at 400~nm (a) and 800~nm (b)
wavelengths at different pump pulse energies from $E_p = 10$~nJ to
100~nJ. An arrow shows a characteristic transmission decrease at
the end of laser pulse. Positive phase corresponds to the $-\Delta
n$ change of refractive index, and vice versa.} \label{f-phase}
\end{center}
\end{figure}

The complex amplitudes at 400~nm and 800~nm were simultaneously
measured using the same optical system. The wide difference of
wavelengths gave a color aberration and the best focusing position
of a lens were different. The deviation was corrected by the
digital focusing using the diffraction calculation based on the
angular spectrum method~\cite{Goodman}. This digital focusing is
one of the advantageous features of the used digital holography
approach.

\section{Results}

In previous study, a side-view imaging at a single 400~nm
wavelength allowed to see formation of strong refractive index
changes, associated with plasma density dynamics, at the earlier
times, followed by shock wave formation, molten flow of material,
and void formation when imaging was carried out till delays of few
nanoseconds~\cite{11ome1399}. Here we focus on the early stages of
plasma formation at smaller excitations.

Figure~\ref{f-ab} shows the amplitude of FFT at two 800 and 400~nm
wavelengths and its axial distribution along the propagation of
the pulse. The amplitude changes are related to the absorptivity
of a plasma at the focal region depending on the energy of the
pump pulse. Slightly stronger amplitude changes observed at 400~nm
wavelengths for the 50~nJ pulse (Fig.~\ref{f-ab}(a)) were related
to a better transmission of the short wavelength pulses through
the plasma region. For the 200~nJ pulse, when a high critical
plasma density is reached even for the shorter wavelength (a
plasma screening), the changes of amplitudes at both wavelengths
were comparable (Fig.~\ref{f-ab}(b)). Subtitle differences of
transmission (and diffraction) through (and around) the plasma in
focal region at two wavelengths was further scrutinized using the
ratio of transmissivity discussed next.

In this new two-color probing, it is informative to present images
as a ratio of the transmitted light at two wavelengths,
$T_{400}/T_{800}$ (Fig.~\ref{f-200}). Transmission at 400~nm has a
larger cut-off (critical) electron density $N_c = \varepsilon_0
m^*\left(\frac{2\pi c}{e\lambda} \right)^2$, where $m^*, e$ are
the optical effective mass and charge of electron, respectively,
$\varepsilon_0$ is the permittivity of vacuum, $c$ is the speed of
light. The blue-color probe (400~nm) can see through the focal
region until the electron plasma density becomes
$N_c^{400~\mathrm{nm}} \simeq 6.8\times 10^{21}$~cm$^{-3}$, while
the 800~nm wavelength light is reflected at the lower plasma
densities when $\sim 1.7\times 10^{21}$~cm$^{-3}$ is reached. A
filament-like low spatial frequency pattern (Fig.~\ref{f-200}) is
recognisable along the pulse propagation; due to the FFT filtering
horizontally extended fringes are apparent in the ratio images
which are not pronounced in the originals (see, Fig.~\ref{f-ab}).
As plasma density is increasing after laser pulse absorption, less
of 400~nm light passes through. The initial stages of absorption
and plasma formation show strong amplitude modulation on the time
scales comparable (or shorter) with pulse duration of 150~fs. Kerr
self-focusing at the very beginning of the transient is
recognisable for the larger pulse energies (also evidenced in
Fig.~\ref{f-phase}).

Transients of the phase changes of FFT images along pulse
propagation reveal real part of the refractive index changes and
are shown in Fig.~\ref{f-phase}. Positive phase changes correspond
to a decrease of the refractive index and vice versa. No permanent
glass modifications were optically recognisable when the pump
pulse energy was smaller than $\sim$20~nJ. Almost twice larger
phase changes were observed for the 400~nm imaging.

The void formation was observed by optical transmission for the
pulses with energy $> 300$~nJ/pulse~\cite{11ome1399}. In the case
of 800~nm probe, there was a fast decrease of phase values, i.e.,
an increase in refractive index, $+\Delta n$, at the end of the
pulse at approximately 250~fs time moment. Thermalisation between
electrons and ions/atoms of the glass matrix has not been finished
at that time and this abrupt change is not related to the matrix
temperature changes. It can be understood as fast removal of free
carriers, i.e., a self-trapping of excitons which leads to defect
formation at the much longer time scales (marked by the arrow in
Fig.~\ref{f-phase}). Self-trapping is a well established in pure
silica glasses where propensity of defect formation after
band-to-band excitation is large. We invoke here the mechanism of
self-trapping of exciton as generic for the wide bandgap
dielectrics (absorption edge of the B270 glass was 300~nm). This
is partly supported by observed similarity of the breakdown and
void formation in the wide silicate glass family which was
distinct from glass forming phosphate and borate
glasses~\cite{07njp253}. The non-bridging oxygen hole center
(NBOHC) is typical paramagnetic defect in silicate glass matrix
which also has luminescence at around 650~nm wavelength ($\sim
1.9$~eV) and is linked to optical damage precursor~\cite{Krol}.
This dynamic process is comparable with pulse duration $\sim
150$~fs. The probe pulse energy is only $E_{pr} = 10$~nJ and its
polarisation is perpendicular to the pump, hence, a coherent cross
talk has to be absent or small due to depolarisation. The phase
contrast after long delays is determined by the mass density
changes and absorbtion bands of defects which can be recalculated
into refractive index changes via the Kramers-Kronig formalism.

In the case of 400~nm probe, there is no fast decrease of the
phase values at the end of the pulse. At this high photon energy a
free carrier absorption can be the major reason for considerably
larger values of the phase ($-\Delta n$); the pulse energy at
400~nm probe was $E_{pr} = 10$~nJ. A recognisable increase in
phase value towards 1~ps and longer times is related to heating of
the matrix which brings a decrease of refractive index, $-\Delta
n$; here we consider that strongly localised heating is expanding
the matrix, hence, local density decrease. Since there is no
apparent reduction of phase values upon formation of self-trapped
exciton under 400~nm probing. It is plausible to assume that
potential barrier for the trapped exciton is larger than 1.5~eV or
800~nm wavelength. It is noteworthy that NBOHC center excitation
has resonant absorption bands and 400~nm wavelength could be out
of the absorbance band.

\begin{figure}[tb]
\begin{center}
\includegraphics[width=8.50cm]{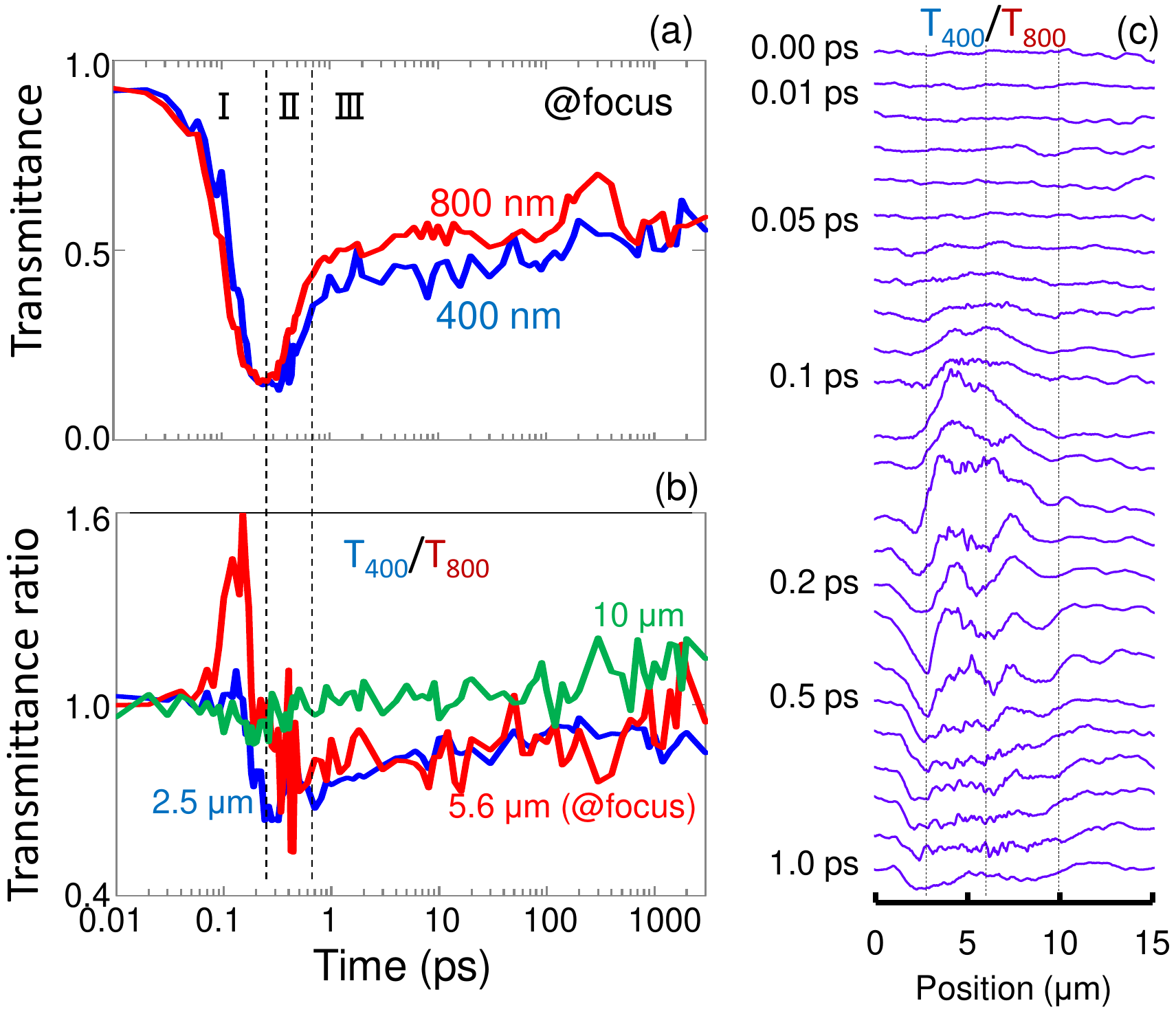}
\caption{(a) Transmittance transient along propagation at 400 and
800~nm at the focal location (or $5.6~\mu$m axial coordinate). (b)
Transmittance ration $T_{400}/T_{800}$ at the different axial
locations before and after focus. (c) Axial cross sections at
different times at focus; central vertical line-marker shows
location of the focus at low pulse energy. Pump pulse energy $E_p
= 200$~nJ.} \label{f-diff}
\end{center}
\end{figure}

Figure~\ref{f-diff} shows the transmissivity ratios at different
locations along propagation of the pulse; the focus is at $z =
5.6~\mu$m. It is possible to define three regions in time
evolution of the differential transmission through the different
locations along pulse propagation. The region I shows processes
occurring during the pulse, the region II is where a fast
evolution of recombination and electron movement before
thermalisation, and the region III captures consequence of
thermalisation between electrons and host matrix.

The largest changes in transmissivity ratio $T_{400}/T_{800}$ are
observed in the region I during the pulse. This is understandable
due to large plasma densities exceeding $10^{20}$~cm$^{-3}$ where
short wavelength light has a higher transmission through the
plasma at the focal region (Fig.~\ref{f-diff}(a)). The
transmissivity ratio is almost unaccented before and after the
focal region along pulse propagation, which indicates that
fillamentation and self-focusing was not strong
(Fig.~\ref{f-diff}(b)). There were on-axis changes in time
observed in the phase and amplitude images in
side-views~\cite{11ome1399}, which are well discernible in
$T_{400}/T_{800}$ ratio (Fig.~\ref{f-diff}(c)). Most probably, it
is related to the plasma density changes during pulse propagation
and self-action of the pulse, i.e., once plasma is excited a back
reflection of light from that region creates a stronger absorption
ahead of the incoming pulse. At the geometrical focus in the
region II, the ratio of transmittances is slightly below one and
has strongest uncertainty. The region III (long delays), a plateau
is observed. In this region light scattering dominates which has
strong size and wavelength dependence $\propto r^6/\lambda^4$ and
affects 400~nm probe significantly (see discussion in the next
section).

\section{Discussion and Outlook}

The proposed method of two-color experiment could be used to
elucidate free electron capture into a trapped states. The
self-trapped excitons are expected to form NBOHC which can reach
high $\sim 10^{19}$~cm$^{-3}$ densities~\cite{16apa194}. The NBOHC
absorbs at 650~nm in silica glasses. Precursors of defects in
glasses and wide bandgap materials~\cite{Krol} can be investigated
by the proposed pump-probe method.

Plasma regions with nanoscale cross sections can be considered as
plasmonic nanoparticles whose extinction cross section - the total
losses measured in transmission  - are due to absorption and
scattering contributions: $\sigma_{ext} \equiv \sigma_{abs} +
\sigma_{sc} \propto r^3/\lambda +  r^6/\lambda^4$ at the
wavelength $\lambda$ for the nanoparticle of radius, $r$. For
example, a 100-nm-diameter gold particle has equal absorption and
scattering cross sections with the latter dominating with
increasing size~\cite{16acsp2184}. Similar scaling is expected for
the metal-like plasma. The size and concentration dynamics of the
plasma regions are expected to bring about complex transients of
transmissivity which are absorption, scattering, and diffraction
dependent. The strong dependence of scattering on the geometrical
size of plasma and wavelength allows to use this method in
investigations of 3D nano-/micro-plasmas important in
high-pressure high-density research~\cite{11nc445}, especially
suitable for a side-view imaging of axially extended focal regions
of Bessel-Gaussian pulses~\cite{Franc,06ol80,01jjap1197}. An
interesting extension of this side-view imaging method can be made
by implementing the four-polarisation method~\cite{Hikima} for
measurement of orientational anisotropy in the sample.

The proposed here high spatial resolution side-view imaging can
find use in stimulated emission depletion STED-inspired fs-laser
fabrication where de-excitation doughnut beam can localise
inter-system crossing into the triplet or defect state down to
nanoscale localisation onto optical axis~\cite{Petit}. The shown
here self-trapping of exciton in glass could potentially be
controlled by STED geometry. This is expected to help a nanoscale
on-demand writing of defects for optical functions, the technique
currently not available for deep-sub-wavelength resolution.

The proposed technique could be applied for two pulses of closely
matching wavelengths (in order to probe the same refractive index)
which are separated enough on the FFT image. This would provide a
direct access to temporal evolution of the refractive index $(n +
ik)$ at the focal region with high temporal resolution and imaging
capability.

\section{Conclusions}

We show a novel two-color probing method suitable to analyse fast
processes induced by a pump pulse in a side-view imaging geometry.
Spatial resolution was set the same by numerical filtering and was
$\sim 0.9~\mu$m for 400 and 800~nm light. Holographic  numerical
refocusing was implemented to compare side view pump-probe images
at different delay times and to use FFT image processing. Fast
decrease of the FFT phase at the end of the pulse for 800~nm
wavelength can be understood as an initial stage of electron
trapping which leads to defect formation at the longer relaxation
times. The transmissivity ratio at different wavelengths allows to
interrogate the opaque regions of plasma, i.e., provides tool to
characterise plasmas of different densities.

\subsection{Acknowledgements}

SJ acknowledges a partial support via the Australian Research
Council Discovery projects DP130101205 and DP170100131. We are
grateful to Prof. Eugene G. Gamaly for discussions on light-matter
interaction at high intensity.


\section*{Author contributions statement}


S.J. and Y.H. conceived the idea of experiments, S-I.F., S.H., and
Y.H. designed experimental setup and tested numerical procedures,
S-I.F. carried out experiments. All the authors participated in
discussion and analysis of the results and contributed to editing
of the manuscript.

\section*{Additional information}

\textbf{Competing financial interests} The authors declare no
competing financial interests.

\end{document}